\documentclass[preprint,pre,a4paper,floatfix,amsmath,longbibliography]{revtex4-1}
\usepackage[utf8]{inputenc}
\usepackage[british]{babel}
\usepackage{graphicx,color}

\usepackage{xr}
\begin{document}

\title{Sharp DNA denaturation in a helicoidal mesoscopic model}
\author{Mateus Rodrigues Leal}
\author{Gerald Weber}

\begin{abstract}
The Peyrard-Bishop DNA model describes the molecular interactions with simple potentials which allow efficient calculations of  melting temperatures. However, it is based on a Hamiltonian that does not consider the helical twist or any other relevant molecular dimensions. Here, we start from a more realistic 3D model and work out several approximations to arrive at a new non-linear 1D Hamiltonian with a twist angle dependence. Our approximations were numerically compared to full 3D calculations, and established its validity in the regime of small angles. For long DNA sequences we obtain sharp, first-order-like melting, transitions.
\end{abstract}

\maketitle

\section{Introduction}

Statistical physics models of DNA using interaction potentials, instead of statistical weights, made a debut with 
\citet{peyrard89}.
This model introduces several simplifications that leaves only a 
single degree of freedom to integrate, transversal to the helical
axis, and for this reason it is commonly referred to as a 1D model~\cite{dauxois91,kalosakas06,buyukdagli10}.
This model allows the calculation of the average base pair displacement, representative of the melting transition, and it was shown that there is an increasing strand opening as temperatures increase.
However, this strand opening occurs only gradually with increasing temperature, which has motivated the search for additional
potentials that could result in much sharper transitions~\cite{dauxois91,peyrard96,weber09}.  
The simplicity of the PB model provides a computational efficiency that can outcompetes atomistic simulations for certain applications, such as describing melting in DNA~\cite{weber06}.
Evidently, the increased efficiency comes at the expense of lack details describing the intramolecular interactions.

In recent years, our group used the mesoscopic Peyrard-Bishop (PB) model for calculating melting temperatures in numerous nucleic acids systems, for instance
for deoxyinosine~\cite{maximiano15}, GU mismatches in RNA~\cite{amarante16} and DNA-RNA hybrids~\cite{martins19}.
Many of our findings correlate well with existing structural data from NMR and X-ray measurements providing
a good level of validation for this theoretical approach.
However, as discussed in some of our previous publications~\cite{weber09b}, the missing helicity and the unusual definition of intramolecular distances of the original PB model~\cite{peyrard89} makes it difficult 
to compare the results with microscopic models, especially to those of molecular dynamics.
The use of an analytical 1D helicoidal Hamiltonian, preferably set in a similar framework as molecular dynamics models~\cite{drukker01}, 
and benefiting from the efficient transfer matrix (TI) technique for calculating the strand separation would be desirable as it may overcome some of the interpretative shortcomings of the PB model.  
 
\begin{figure}[htbp]
    \centering
    \includegraphics[width=0.8\columnwidth]{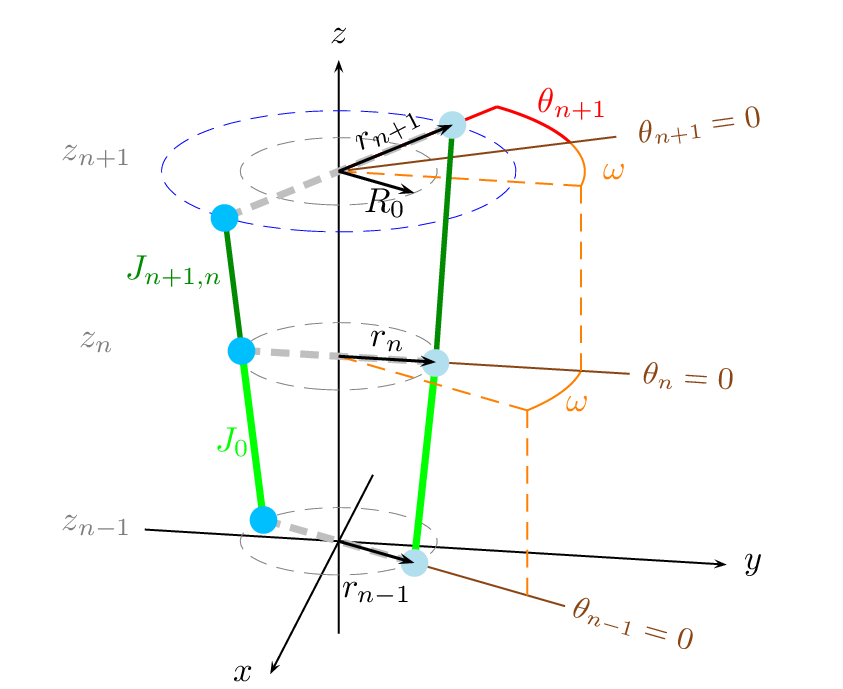}
\caption{Schematic diagram of the 3D model.
Base pairs $n-1$ and $n$ are shown at their equilibrium positions, and the base pair $n+1$ is rotated by $\theta_{n+1}$ and stretched from its equilibrium position to $r_{n+1}$.
The twist angles $\omega$ (orange lines) defines the offset between consecutive base pairs angle origins~$\theta=0$ (brown lines).
Green lines show the stacking interaction between neighbouring bases, in particular the distances $J_{n+1,n}$ are shown as a dark green lines and $J_0$ are the equilibrium stacking distance shown as light green lines.
The grey circles show the radius of the equilibrium distance $R_0$ and the blue circle shows the stretched distance $r_{n+1}$.
Hydrogen bond distances are shown as thick dashed grey lines.
\label{fig-3d-h}\label{fig-3d}}
\end{figure}

There have been several proposals for helicoidal Hamiltonians within the framework of TI partition function calculation of the PB model~\cite{cocco99,barbi99,cocco00}.
These models add the helical twist angle, but also add constraints that make it difficult to integrate analytically the
partition function.
The common approach is to fix beforehand the distance $z_{n+1}-z_{n}$ between consecutive base pairs, see fig.~\ref{fig-3d-h}, known as the helical rise distance $h$~\cite{olson01}.
By fixing $h$, the radial distance~$r$ and the twist angle~$\theta$ both need to be integrated numerically~\cite{cocco99,barbi03,michoel06}.
Unlike the PB model, the integration of the helicoidal Hamiltonian, restricted in this way, does not result in an analytical 1D Hamiltonian. 
Other helicoidal models based on the PB Hamiltonian that do not calculate melting transitions or do not use the transfer integral method, such as refs.~\citenum{tabi09,behnia11b,torrellas12,zoli18,zdravkovic19,nomidis19}, are not covered here. 

Our aim is to adapt a 3D Hamiltonian, with distances and angles as shown in fig.~\ref{fig-3d-h}, and obtain an analytical expression for a 1D Hamiltonian with included twist angle dependence.
A constant twist angle~$\omega$ defines the origin of the angles~$\theta$ at each base pair.
The setup of fig.~\ref{fig-3d-h} follows closely that of B-DNA, which from crystallographic measurements it is known to  
have a tilt angle of -0.1$^\circ$ and roll of 0.6$^\circ$ per base pair step~\cite{olson01}.
In other words, for short DNA sequences there is no appreciable bending and the configuration shown in fig.~\ref{fig-3d-h} is justified.

We integrate the 3D partition function by carefully introducing approximations and restrictions and arrive at a new 1D Hamiltonian with radial distance and twist angle dependence.
For small twist angles, the resulting melting transitions are sharp first-order-like with strong discontinuity of the strand opening for very long DNA sequences.
On the other hand, as soon as the twist angles are increased these transitions rapidly loose their strength. 
To evaluate the impact of the approximations, we numerically integrated the configurational part of the partition function of the 3D Hamiltonian.
We repeat these numerical integrations also applying similar restrictions that were used for the 1D Hamiltonian.
These numerical tests show that the results from the helicoidal 1D Hamiltonian are qualitatively similar to the full 3D model within the regime of small angles.

\section{Model}

The configurational part of the classical partition of a oligonucleotide duplex composed of $N$ base-pairs 
is written in terms of the polar cylindrical coordinates $z$, $r$ and $\theta$
\begin{equation}
Z_{r\theta z} = \Gamma^N \int
\prod_{n=1}^{N}dz_nd\theta_nr_ndr_n 
\exp\left[-\beta U_{n,n-1}(z_n,\theta_n,r_n,z_{n-1},\theta_{n-1},r_{n-1})\right]
\label{eq-Z}
\end{equation}
where $\beta=1/(k_BT)$, $k_B$ is the Boltzmann constant and $T$ the absolute temperature.
$\Gamma$ is a density factor, which is taken here as a reciprocal unit of volume, such that $Z_{r\theta z}$ becomes adimensional.
$U$ is the configurational part
of model Hamiltonian and is a function of the ($z,\theta,r$) positions of two consecutive base pairs.
The customary periodic boundary condition, where the last base-pair interacts with the first, is represented by the potential $U_{1,N}$.
The average radius~$\langle r_k\rangle$, representing the intra-strand opening, can be calculated as follows
\begin{equation}
\langle r_k\rangle = \frac{\Gamma^N}{Z_{r\theta z}} \int
\prod_{n=1}^{N}r_kdz_nd\theta_nr_ndr_n 
\exp\left[-\beta U_{n,n-1}(z_n,\theta_n,r_n,z_{n-1},\theta_{n-1},r_{n-1})\right]
\label{eq-r-av}
\end{equation}
For the case where all model parameters are the same at each site~$k$, we have 
$\langle r\rangle=\langle r_1\rangle=\ldots=\langle r_N\rangle$.

The origins of the $\theta$~angles of consecutive base pairs are offset by a fixed twist angle~$\omega$ between base pair steps, see fig.~\ref{fig-3d-h},  which allows the use of an single integration limit for all angle variables, that is, $\theta_n \in [-\Theta,\Theta]$.
For $r_n$ the integrations are taken within the limits $r_n \in [0,b]$.
For $z$ we integrate within a region $\pm\zeta$ around the rise distance~$h_0$ such that the limit is taken as 
\begin{equation}
z_n \in [(n-1)h_0-\zeta,(n-1)h_0+\zeta]
\end{equation}
the partition function is then written with explicit integration limits as
\begin{eqnarray}
Z_{r\theta z} 
&=&
\Gamma^N \int_{0}^{b} \int_{-\Theta}^{\Theta} \int_{(n-1)h_0-\zeta}^{(n-1)h_0+\zeta}
\prod_{n=1}^{N}dz_nd\theta_nr_ndr_n 
\exp\left[-\beta U_{n,n-1}(z_n,\theta_n,r_n,z_{n-1},\theta_{n-1},r_{n-1})\right]
\label{eq-Z-limits}
\end{eqnarray}
where each integration symbol implies $N$-uple integrals. 
The interaction potential $U$ is divided into stacking interactions $W_{n,n-1}$ and base-pair interactions $V_n$,
\begin{eqnarray}
&&U_{n,n-1}(z_n,\theta_n,r_n,z_{n-1},\theta_{n-1},r_{n-1})=
V_n(r_n) 
+ W_{n,n-1}(z_n,\theta_n,r_n,z_{n-1},\theta_{n-1},r_{n-1})
\end{eqnarray}

In terms of the polar cylindrical coordinates and considering the 3D scheme shown in fig.~\ref{fig-3d}, the base-pair
interaction potential is solely a function of $r_n$, that is $V_n(r_n)$ and it brings no difficulty for the
integration of eq.~(\ref{eq-Z-limits}).
Here we will use the Morse potential
\begin{equation}
V_n(r_n)=D\left[e^{-a(r-R_0)}-1\right]^2
\label{eq-V}
\end{equation}
where $D$ is the depth of the potential, $a$ the width and $R_0$ an equilibrium distance.
The stacking interaction~$W$ however depends on all coordinates and links to consecutive base-pairs $n$ and $n+1$, which
is the main point of difficulty for a full algebraic integration of the partition function, eq.~(\ref{eq-Z-limits}).
Therefore, our efforts will centre on the handling of the 3D stacking potential and, unlike the base-pair potential, 
the specific form of this potential is a crucial aspect of the theoretical method.
Here, the stacking interaction potential is given by the harmonic potential between neighbouring bases $n$ and $n-1$,
\begin{equation}
W_{n,n-1} =  \frac{k}{2} \left(J_{n,n-1}-J_0\right)^2,
\end{equation}
where $J_{n,n-1}$, shown as a green line in fig.~\ref{fig-3d}, is the distance between two bases belonging to the same strand.
$J_0$ is the equilibrium distance and $k$ the elastic constant.
In polar cylindrical coordinates $z$, $r$ and $\theta$, shown in fig.~\ref{fig-3d}, the distance $J_{n,n-1}$ is written as
\begin{equation}
J_{n,n-1}=\sqrt{\Delta z_{n,n-1}^2+f_{n,n-1}^2}
\label{eq-Jnn}
\end{equation}
where $f_{n,n-1}$ is the $xy$-projection 
\begin{equation}
f_{n,n-1} = \sqrt{r_{n}^2+r_{n-1}^2 -2r_{n}r_{n-1}\cos(\omega +\theta_n-\theta_{n-1})}
\end{equation}

The equilibrium distance between the two consecutive base-pairs along the $z$-axis is~$h_0$, corresponding to the rise
distance and $\omega$ is the structural twist angle~\cite{olson01}.  
For simplicity, we will assume that both bases at the $n$th site are at the same distance in regard to the~$z$ axis, that is, 
they move symmetrically with respect to the helical $z$~axis.
While this may seem overly restrictive, we have shown that for the classical partition function in the PB~model this means that
the elastic constant is simply the average of the elastic parameters of each strand~\cite{martins17}.
Therefore, the elastic constant $k$ is the equivalent constant of the two springs to each side of the duplex strand.

We now expand eq.~(\ref{eq-Jnn}) to first order of $\Delta z_{n,n-1}^2$
\begin{equation}
J_{n,n-1} \approx \Delta z_{n,n-1}
\left[
1+\frac{1}{2}\frac{f_{n,n-1}^2}{\Delta z_{n,n-1}^2}
\right].
\label{eq-J-approx}
\end{equation}
To higher orders of $\Delta z_{n,n-1}^2$ the remaining equations become quite complicated.
Therefore, for the sake of the discussion, we will present here only the simpler development following the first order expansion
without loss of generality, and show the more complicated expansion to second order in supplementary equations 
Eqs.~(\ref{supp-eq-J-approx2}--\ref{supp-eq-Z-approx}).
We now use the additional restriction 
\begin{equation}
\Delta z_{n,n-1}\approx J_0
\label{eq-zJ-rest}
\end{equation}
which is similar as used by other authors~\cite{cocco99,barbi03,michoel06}.
However, the crucial difference here is that we apply
it after the expansion of eq.~(\ref{eq-J-approx}), as it enables us to carry out the remaining integrations and arrive at an analytical form for the 1D Hamiltonian, which is the
aim of this work.
After integration in $z$, and the
partition function simplifies to
\begin{eqnarray}
Z_{r\theta z} &=& \Gamma (2\zeta)^N
\int_{0}^{b} \int_{-\Theta}^{\Theta}
\prod_{n=1}^{N}dr_nd\theta_nr_n 
e^{-\beta V(r_n)}
\exp\left(-\beta k \frac{f_{n,n-1}^4}{8J_0^2}\right)
\label{Z-app-z}
\end{eqnarray}
For the angle integration we will use $\theta_n-\theta_{n-1} \ll \omega$, and the approximation 
\begin{equation}
\cos(\omega +\theta_n-\theta_{n-1})\approx 1-\frac{\omega^2}{2}
\label{eq-cos-rest}
\end{equation}
Note that a small difference $\theta_n-\theta_{n-1}$ does not imply in a flattened helix, since the angles are always offset
by the helical twist $\omega$, see fig.~\ref{fig-3d}.
Integrating over $\theta$, we arrive at the final approximated form of the partition, after rearranging terms to symmetrize
the integrand 
function
\begin{eqnarray}
Z^{app.}_{r\theta z} &=& \Gamma^N (4\zeta \Theta )^N
\int_{0}^{b} 
\prod_{n=1}^{N}dr_n
\sqrt{r_nr_{n-1}} 
\exp\left\{-\frac{\beta}{2}\left[ V(r_n)+V(r_{n-1})\right]\right\}
\nonumber \\ &&\times
\,\exp\left\{-\frac{\beta k}{8J_0^2}\left[\left(r_n-r_{n-1}\right)^2 +\omega^2r_nr_{n-1}\right]^2 \right\}
\label{eq-Z-approx}
\end{eqnarray}
Note that the fluctuations along and around the $z$-axis are given by $\zeta$ and $\Theta$, respectively, which are now outside the remaining
integration, therefore those factors will simply cancel out when calculating expectation values, eq.~(\ref{eq-r-av}).

The remaining variable to integrate is in $r_n$ which can be handled by the transfer integral technique
where the kernel is
\begin{equation}
K(x,y)
=
(xy)^{1/2}e^{-\frac{\beta}{2}\left[ V(x)+V(y)\right]}
\,\exp\left\{-\frac{\beta k}{8J_0^2}\left[\left(x-y\right)^2 +\omega^2xy\right]^2 \right\}
\label{eq-kernel-3D}
\end{equation}
In effect, this is now equivalent to a one-dimensional radial Hamiltonian with a twist angle dependence
\begin{equation}
U(r_n,r_{n-1})
= V(r_n)
+\frac{k}{8J_0^2}\left[\left(r_n-r_{n-1}\right)^2 +\omega^2r_nr_{n-1}\right]^2
\label{eq-ham-helic}
\end{equation}

The approximated partition function eq.~(\ref{eq-Z-approx})
can be evaluated via the transfer integral (TI) technique~\cite{peyrard89,zhang97b}.
In this technique the kernel is discretized over $M$ points in the interval $[0,b]$ and the partition function becomes
\begin{equation}
Z_{TI}=\Gamma^N (4\zeta \Theta )^2\sum_{i=1}^M \lambda_i^N
\label{eq-z-ti}
\end{equation}
The average radius $\langle r\rangle_{TI}$ is calculated as
\begin{equation}
\langle r\rangle_{TI}=\frac{\sum_{i=1}^M \lambda_i^N\int_0^b |\phi_i|^2 dr}{\sum_{i=1}^M \lambda_i^N}
\label{eq-r-ti}
\end{equation}
where $\phi_i$ are the eigenfunction.
For details of this procedure see Refs.~\citenum{peyrard89,zhang97b,weber09}.
For the limit $N\rightarrow\infty$ this further simplifies to
\begin{equation}
\lim_{N\rightarrow\infty}\langle r\rangle_{TI}=\int_0^b |\phi_1|^2 dr
\label{eq-lim}
\end{equation}
where $\phi_1$ is the eigenfunction with the highest eigenvalue $\lambda_1$~\cite{peyrard89}.
We will refer to the approximation calculated by the TI technique as T1 and T2, for the first and second order expansion
of eq.~(\ref{eq-Jnn}), kernels eqs.~(\ref{eq-kernel-3D}) and (\ref{supp-eq-kernel-3D}), respectively.

\subsection{Numerical tests}

Here, we will compare numerically the approximated eq.~(\ref{eq-z-ti}) to the fully integrated partition function eq.~(\ref{eq-Z-limits}).
To our knowledge, the numerical evaluation of the 3D Hamiltonian, eq.~(\ref{eq-Z-limits}).
One possible reason for this is that the numerical effort scales with $N^3$,
even for the smallest possible number of base pairs, $N=2$, this is very much on the limit of computational feasibility.
For $N=2$ the numerical integration has taken us of the order of days, even with parallel processing.
Therefore, we are limited to $N=2$ for the evaluation of eq.~(\ref{eq-Z-limits}).
On the other hand, the TI solution eq.~(\ref{eq-z-ti}) is valid for sequences of any length~$N$.

We will keep the periodic boundary condition, which may seem odd for a sequence of length of just $N=2$, however there is no
loss of generality for the results presented here.
The reason for this is that a sequence of length $N=2$ has two elastic constants, $k_{1,2}=k_{2,1}=k$, where the last one represents the periodic boundary condition.
The open boundary condition is simply setting $k_{1,2}=k$ and $k_{2,1}=0$~\cite{zhang97b}, which for $N=2$ turns out to be the exact equivalent
of maintaining the periodic boundary condition and setting $k_{1,2}=k_{2,1}=k/2$.

We designated the partition function calculated from  eq.~(\ref{eq-Z-limits}) as $Z_{C}$, where C stands for complete,
\begin{equation}
Z_{C}=Z_{r\theta z}\left[N=2,b,\zeta,\Theta\right]
\label{eq-Z-N}
\end{equation}
Furthermore, we calculate eq.~(\ref{eq-Z-N}) by adding the restrictions of eqs.~(\ref{eq-zJ-rest}) and (\ref{eq-cos-rest}),
which we called the restricted (R) calculation, which is a subset of the $Z_{C}$ calculation,
\begin{equation}
Z_{R}=Z_{C}\left[\Delta z_{n,n-1}\approx J_0;\cos(\omega +\theta_n-\theta_{n-1})\approx 1-\frac{\omega^2}{2}\right]
\end{equation}
and it is expected that the TI calculations should be close to~R. 
Details of the numerical integrations are given in supplementary section~\ref{supp-sec-int}.

Unless noted otherwise, for the numerical tests we used the following parameters: $D=0.2$~eV, $a=42.5$~nm$^{-1}$, $k=4$~eV/nm$^{2}$, $J_0=0.7$~nm,
corresponding to a homogeneous oligonucleotide sequence, and were largely chosen to highlight the main
differences in the integration methods.
The value for $J_0$ was adapted from Ref.~\citenum{cocco99}.
The equilibrium distance was taken as $R_0=0.1$~nm, as $r$ represents half the distance between the base pairs,
this corresponds to a hydrogen equilibrium bond distance of $2R_0=0.2$~nm.

\section{Results and discussion}

\begin{figure*}[tbp]
\begin{center}
\includegraphics[width=0.8\textwidth]{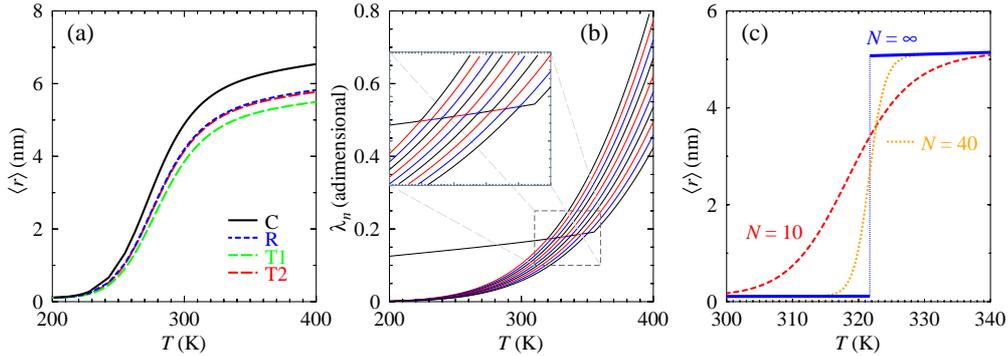}
\end{center}
\caption{Panel (a): calculated average radius~$\langle r\rangle$ as function of temperature for calculations of type C (black curve), R (blue short dashed), T1 (green long dashed) and T2 (red short dashed).
Panel~(b): 10 highest eigenvalues~$\lambda_n$ for T1 (T2 in fig.~\ref{supp-fig-zr-t}), the inset shows a zoom-in to highlight the anti-crossings.
Panel~(c): average radius~$\langle r\rangle$ for sequence length $N=10$ (red dashed), $40$ (orange dotted) and $\infty$ (thick blue), the vertical dotted blue line shows the discontinuity for $N\rightarrow\infty$ within resolution of $\Delta T=0.01$~K.    
Integration limits are $\zeta = 10^{-2}$~nm, $\omega=0.05$~rad, $\Theta = 0.01$~rad and~$b=20$~nm.
\label{fig-zr-t}}
\end{figure*}

The dependence of the average radius $\langle r\rangle$ as function of temperature is shown in fig.~\ref{fig-zr-t}a for the numerical tests C, R, T1 and T2. 
In all cases, the denaturation curves exhibit the characteristic sigmoidal shape of the melting transition that has been
the characteristic of the Peyrard-Bishop model~\cite{peyrard89}.
The approximated calculation to first order expansion, T1, underestimates the average radius when compared to the C and R calculations,
especially as temperature increases.
The restrictions of eqs.~(\ref{eq-zJ-rest}) and (\ref{eq-cos-rest}) do represent a substantial part of this reduction, as shown
by the differences between the C and R calculations.
This is to be expected as all three approximations, eq.~(\ref{eq-J-approx},\ref{eq-zJ-rest},\ref{eq-cos-rest}), essentially limit
the scope of the integration thus resulting in smaller $\langle r\rangle$.
However, when we use the second order expansion T2, the result is very close to the R calculation, therefore the differences
between T1 and R are only due to the order of the expansion of eq.~(\ref{eq-Jnn}).
The spectrum of $\lambda_n$ for T1 is shown in fig.~\ref{fig-zr-t}b (see fig.~\ref{supp-fig-zr-t} for T2) and displays the characteristic anti-crossing between successive
eigenvalues~\cite{peyrard96}, which is highlighted in the zoomed-in inset.
Unlike the spectra of the 1D models~\cite{weber06b} where the eigenvalues have a substantial gap at the anti-crossings, here in fig.~\ref{fig-zr-t}b
this gap is barely noticeable.
As the sequence length $N$ increases the transition becomes increasingly abrupt, as shown in fig.~\ref{fig-zr-t}c.
In the limit of $N\rightarrow\infty$, see eq.~(\ref{eq-lim}), a discontinuity is observed for $\Delta T=0.01$~K, similar to what was found by \citet{barbi03}.

\begin{figure*}[tbp]
\begin{center}
\includegraphics[width=0.6\textwidth]{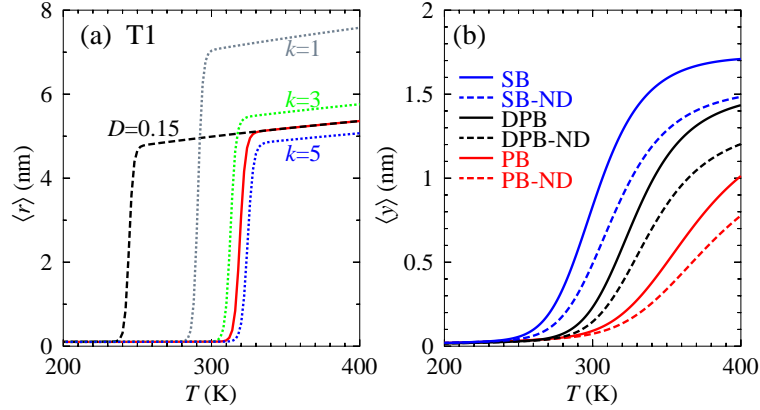}
\end{center}
\caption{Calculated average radius~$\langle r\rangle$ for T1 (panel~a, for T2 see fig.~\ref{supp-fig-r-n}) and calculated 1D average displacement~$\langle y\rangle$ (panel~b) as function of temperature for sequences of size $N=25$. 
Panel (a): Morse parameters parameters are $D=0.2$~eV for all curves, except dashed black curve for which  $D=0.15$~eV was used;
Stacking parameters are $k=4$~eV/nm$^2$ for the red solid curve and black dashed curve, the remaining dotted curves are indicated in eV/nm$^2$. 
Parameters for the T1 were $\zeta = 10^{-2}$~nm, $\omega=0.05$~rad, $\Theta = 0.01$~rad and~$b=20$~nm. 
Panel~(b): parameters for the 1D DPB model as in ref.~\citenum{dauxois93}, and for PB and SB as in ref.~\citenum{weber06b}.
Also shown as dashed curves, with suffix ND (non-divergent), are the calculations with added angle $\omega=0.01$ 
as in eq.~(\ref{eq-K2d-cos}).
\label{fig-r-n}}
\end{figure*}

For moderate sequence lengths, for instance $N=25$, the helicoidal model already shows much steeper transitions (fig.~\ref{fig-r-n}a) than other PB-type models, fig.~\ref{fig-r-n}b.
Some examples of different model parameters are shown in fig.~\ref{fig-r-n}a.
Varying the Morse potential~$D$ changes the temperature where the transitions occurs but not the $\langle r\rangle$ at high temperatures.
The stacking parameter~$k$ on the other hand has an influence on both the onset of the transition and the high temperature value of $\langle r\rangle$.

To understand the differences between the helicoidal and PB-type models it is instructive to look at the 
symmetrized kernel eq.~(\ref{eq-kernel-3D})  used for the T1 calculation, see supplementary eq.~(\ref{supp-eq-kernel-3D}) for T2,
and compare to the PB kernel~\cite{peyrard89}
\begin{equation}
K_{PB}(x,y)=e^{-\frac{\beta}{2}\left[V(x)+V(y)\right]}
\exp\left[-\beta k \frac{ (x-y)^2}{2}\right]
\label{eq-K2d}
\end{equation}
One important difference is the $(x-y)$ to the fourth order in stacking term of eq.~(\ref{eq-kernel-3D}), instead of second order for the PB kernel~\cite{peyrard89}.
Therefore, the harmonic 3D stacking seemingly maps into an anharmonic stacking term in the helicoidal model.
However, in our tests with the helicoidal model, such a fourth power term is not the main cause of a steep transition (data not shown), although
it has an important influence on which temperature the transition starts and how large $\langle r\rangle$ becomes at higher
temperatures. 
What actually ensures the abrupt rise of $\langle r\rangle$ is the $(xy)^{1/2}$ factor which comes from the integration in~$r$, which
does not exists in the PB~model.
It also has an effect on magnitude of the potential parameters.
For instance, the $D$ we used to obtain a transition at higher temperatures is much closer actual energies of the hydrogen bonds, between 0.15 and 0.4~eV~\cite{drukker01,wendler10} whereas for the PB model these potentials are typically an order of magnitude smaller~\cite{weber09b}.
The last factor in eq.~(\ref{eq-kernel-3D}) contains the twist angle~$\omega$ which plays a crucial role in preventing the divergence
in the integration, we will discuss this in more detail next.

\begin{figure*}[tbp]
\begin{center}
\includegraphics[width=0.8\textwidth]{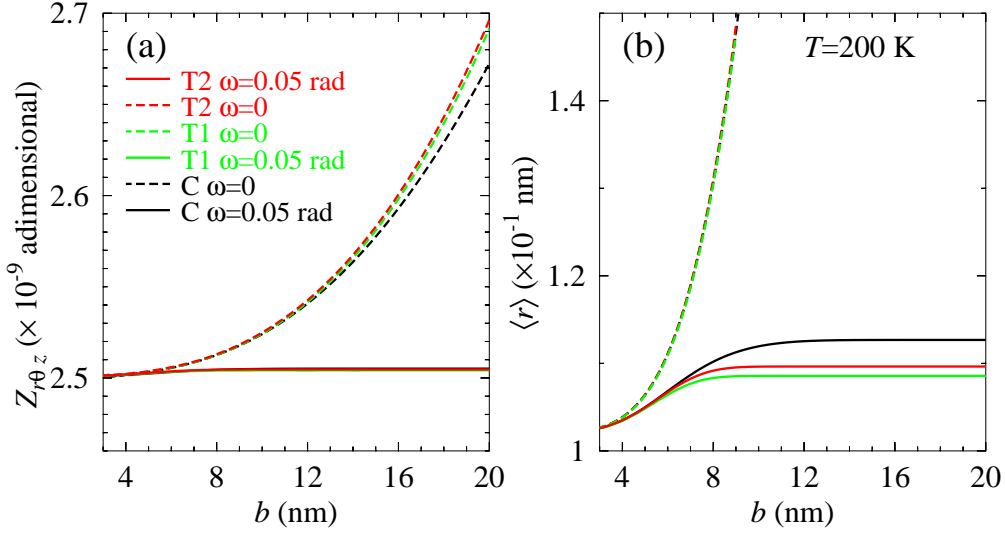}
\end{center}
\caption{Calculated (a)~configurational part of the partition function~$Z_{r\theta z}$ and (b)~average radius~$\langle r\rangle$ as function of the upper limit~$b$ of the integration variable~$r$, at 200~K, for C (black), T1 (green) and T2 (red).
Full curves are for a twist angle of $\omega=0.05$~rad and dashed curves for $\omega=0$.
Integration limits are $\zeta = 10^{-2}$~nm,  $\Theta = 0.01$~rad.
\label{fig-zr-b}}
\end{figure*}

All PB-type models suffer from a numerical divergence, this is becomes especially apparent for the anharmonic Dauxois-Peyrard-Bishop (DPB) model
and was discussed in detail by \citet{zhang97b}.
One tentative approach to circumvent this divergence was to add a twist angle $\omega$ to eq.~(\ref{eq-K2d})~\cite{weber06},
\begin{equation}
(x-y)^2\rightarrow(x^2-2xy\cos\omega +y^2)
\label{eq-K2d-cos}
\end{equation}
which mimics a small out of plane angle.
This procedure avoids the divergence for any PB model~\cite{weber06b} but also reduces the steepness of the transition,
see dashed curves fig.~\ref{fig-r-n}b.
In general, the solvent-barrier (SB) model~\cite{weber06b}, another PB-type Hamiltonian, has a much steeper increase of the displacement than the anharmonic DBP~\cite{dauxois93} or the original harmonic PB model~\cite{peyrard89}.
The helicoidal model also shows the divergence problem if the twist angle is zero, $\omega=0$, as shown
in fig.~\ref{fig-zr-b}.
The radius~$\langle r\rangle$ diverges much more strongly than the partition function~$Z_{r\theta z}$ 
due to the additional variable $r$ in the integration of eq.~(\ref{eq-r-av}).
Therefore the onset of the divergence for $\langle r\rangle$, fig.~\ref{fig-zr-b}b, occurs at a much shorter~$b$ than for  
$Z_{r\theta z}$, fig.~\ref{fig-zr-b}a.
The divergence appears equally for the C and TI calculations, and consequently is not a particularity introduced by the
approximations or by the transfer integral technique.
Setting the twist angle $\omega$ to a non-zero value, no matter how small, removes the divergence entirely and
therefore brings some justification to the similar approach used in the PB model, eq.~(\ref{eq-K2d-cos})~\cite{weber06b}.

\begin{figure*}[tbp]
\begin{center}
\includegraphics[width=0.8\textwidth]{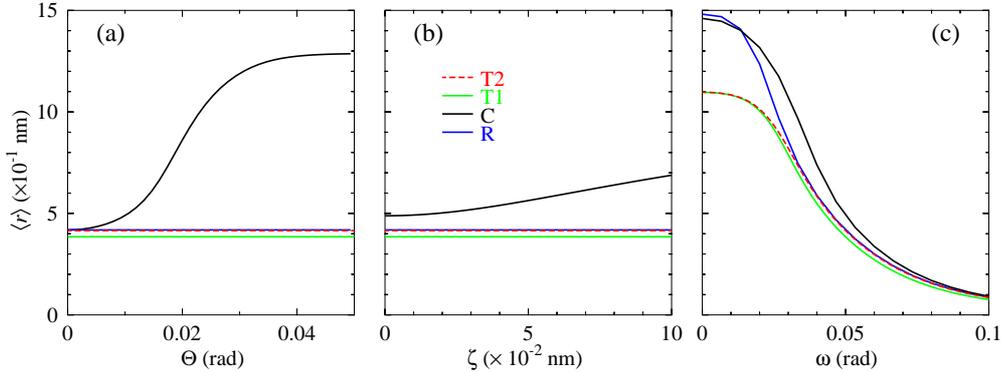}
\end{center}
\caption{Average radius~$\langle r\rangle$ as function of the upper limit (a)~$\Theta$, (b)~$\zeta$ and (c)~$\omega$, at temperature 300~K
and $\omega=0.05$~rad; for C (black), R (blue), T1 (green) and T2 (red).
Integration limits are (a,b)~$b=20$~nm , (c)~$b=15$~nm , and (a)~$\zeta = 10^{-2}$~nm or (b)~$\Theta = 0.01$~rad.
\label{fig-r-theta-zeta}}
\end{figure*}

The average radius~$\langle r\rangle$ follows in general that of the restricted numerical integration~R.
Deviations of T1, T2 and R, from the unrestricted calculation~C, become larger if we move away from the
conditions where the approximations are valid, which is for small angles~$\Theta$ and small longitudinal
displacements~$\zeta$, see fig.~\ref{fig-r-theta-zeta}a,b. 
The limit~$\Theta$ of the angle~$\theta$ and the upper limit~$\zeta$ for the $z$~variable both appear as
constant factors in eq.~(\ref{eq-Z-approx}) and therefore are cancelled in the calculation of the average radius. 
As a consequence, the average radius is constant for $\theta$ and $\zeta$, 
for the approximated calculation as shown in fig.~\ref{fig-r-theta-zeta}a,b.
The same happens for~R, which validates the T1 and T2 within these restriction.
For the twist angle~$\omega$ we observe a progressive reduction of the average radius~$\langle r\rangle$ after 
$\omega=0.02$.
For larger angles, $\langle r\rangle$ tends towards the equilibrium radius~$r_0$ in all cases, which is consistent with the
idea that the strands can not separate without unwinding the helix.

\section{Conclusions}

We have developed and tested several approximations that allow the 3D Hamiltonian to be analytically integrated 
and resulted in a new 1D Hamiltonian with twist angle dependence. 
A first-order-like transition is observed, much steeper than for any PB model.
This transition arises naturally in the helicoidal model, without the need of additional anharmonic potentials.

The results of the new helicoidal model, when compared to the restricted and unrestricted 3D calculations, points to a qualitative 
agreement in regime of small angles.
Therefore, this approximated model, in particular represented by the Hamiltonian of eq.~(\ref{eq-ham-helic}), is expected 
to be useful for situations where the DNA helix is completely unwound.
This is typically the case close to the temperature of DNA denaturation.
We believe that its primary use will be for replacing PB-like Hamiltonians in melting temperature calculations~\cite{weber06},
as it can be used within the framework of the TI method that already exists for the PB models~\cite{weber13b}.
The helicoidal model considers a similar structural definition as used in molecular dynamics~\cite{drukker01}, which
enables the use of compatible parameters, such as hydrogen bond equilibrium distances.
In addition, we showed that the Morse potential parameters are now much closer to those that are obtained from
quantum mechanical calculations~\cite{wendler10}. 

\section{Supplementary Information}

Supplementary equations~(\ref{supp-eq-J-approx2}--\ref{supp-eq-kernel-3D}) are the T2 algebraic development. 
Supplementary figure~\ref{supp-fig-zr-t} is the T2 equivalent of figure~\ref{fig-zr-t}, figure~\ref{supp-fig-r-n} the T2 equivalent of~figure~\ref{fig-r-n}.
Supplementary section \ref{supp-sec-int} describes the details of the numerical integration.

\section{Acknowledgements}

This work was supported by Funda\c{c}\~ao de Amparo a Pesquisa do Estado de Minas Gerais (Fapemig);
Conselho Nacional de Desenvolvimento Científico e Tecnológico (CNPq); and 
Coordena\c{c}\~{a}o de Aperfei\c{c}oamento de Pessoal de N\'{i}vel Superior (CAPES). 

\section*{References}
\bibliography{complete-gbc}

\end{document}